\documentclass[aps,pra,showpacs,twocolumn]{revtex4}
\usepackage{amsfonts}

\usepackage{amssymb}
\usepackage{amsmath}
\usepackage{graphicx}
\usepackage{epsfig}

\input{tcilatex}

\begin{document}

\title{Vanishing environment-induced decoherence }

\begin{abstract}
For a central system uniformly coupled to a XY spin-1/2 bath in a transverse
field, we explicitly calculate the Loschmidt echo(LE) to characterize
decoherence quantitatively. We find that the anisotropy parameter $\gamma $
affects decoherence of the central system sensitively when $\gamma \in [0,1]$%
, in particular, the LE becomes unit without varying with time when
$\gamma =0 $ , implying that environment-induced decoherence
vanishes . Some other cases in which the LE is unit are discussed
also.
\end{abstract}

\author{Yong-Cheng Ou and Heng Fan}
\affiliation{Institute of Physics, Chinese Academy of Sciences, Beijing 100080, China}
\maketitle

\section{Introduction}

Coherence of a quantum state is very fragile because of the existence of \
its environmental degrees of freedom coupled to it, which has become the
major obstacle in constructing quantum computer \cite{AGG,A1GG}. To protect
the quantum information, we generally use the quantum error correction
scheme which can correct the quantum errors to protect the encoded quantum
states\cite{CS,S,Gottesman}; We can also use the scheme to find the
decoherence-free subspaces and some other schemes to protect the quantum
information\cite{Duan,Lidar}. Generally we need several physical qubits to
realize one logic qubit in these schemes. It will be very interesing if we
can find a quantum systems in which the quantum states can be naturally
protected.

On the other hand, many physicists took attention to the
relationship among the concepts of environment, decoherence, and
irreversibility, these investigations may provide new perspective of
how to overcome decoherence and renewed understanding for the
crossover between quantum and classical behavior \cite{zurek}. In
the study of quantum-classical transition in quantum chaos, the
concept of Loschmidt echo(LE) was introduced\cite{peres}, and we
will use it to determine decoherence of a central system.

With the development of quantum information, the concept of entanglement
(concurrence) was used to investigate the quantum phase transition(QPT) \cite%
{osborne,vidal}. Recently, Carollo and Pachos \cite{carollo} have
established the relation between geometric phases(GP) and
criticality of spin chain, Zhu \cite{zhu} analyzed the scaling of
the geometric phases, and Hamma \cite{hamma} found that a
non-contractible GP of the ground state is also a witness of QPT.
Quan \textit{et al} \cite{quan} found that the decay of the LE is
enhanced by quantum criticality in Ising spin-1/2 model. These
particular features in the XY model above seem to imply that there
exist distinct properties with respect to decoherence of a central
quantum system surrounded by it. In this paper, we show that the
anisotropy parameter $\gamma $ affects decoherence of the central
system sensitively when $\gamma \in [0,1]$, in particular, the LE
becomes unit without varying with time when $\gamma =0 $ , implying
that environment-induced decoherence vanishes . Some other cases in
which the LE is unit are discussed also.

\section{Derivation of the Loschmidt echo for a central system}

Firstly, we analyze the XY model as a starting point, since it is
exactly solvable and presents a rich structure. The system-bath
model can be
described by the Hamiltonian $\mathcal{H}=\mathcal{H}_{S}+\mathcal{H}_{E}+%
\mathcal{H}_{SE}$, the central (two-state) system, characterized by the
ground state $\left\vert 0\right\rangle $\ and the excited state $\left\vert
1\right\rangle ,$ has a free Hamiltonian $\mathcal{H}_{S}=w_{e}\left\vert
1\right\rangle \left\langle 1\right\vert $ and is coupled to all spins in
the bath through the interaction $\mathcal{H}_{SE}=-\delta
\sum\nolimits_{l}\left\vert 1\right\rangle \left\langle 1\right\vert \sigma
_{l}^{z},$ where $\delta $ represents the coupling constant. Our model
differs from the model in \cite{rossini} where the system only interacts
with the first spin in the bath. The chain in a transverse field has nearest
neighbor interactions with Hamiltonian expressed by
\begin{equation}
\mathcal{H}_{E}=-\sum\limits_{l=-M}^{M}\left( \frac{1+\gamma }{2}\sigma
_{l}^{x}\sigma _{l+1}^{x}+\frac{1-\gamma }{2}\sigma _{l}^{y}\sigma
_{l+1}^{y}+\lambda \sigma _{l}^{z}\right) ,  \label{1}
\end{equation}%
where $M=(N-1)/2$ for $N$ odd, and the operators $\sigma _{l}^{\alpha }$ ($%
\alpha =x,y,z$) are the usual Pauli operators defined on the $l$-th site of
the lattice. The constants $\gamma \in \lbrack 0,+\infty ]$ and $\lambda \in
\mathbb{R}$ represent the anisotropy parameter in the next-neighbor
spin-spin interaction and an external magnetic field. The model defined by
Eq.(\ref{1}) has a rich structure \cite{book}, i.e., when the anisotropy
parameter $\gamma $ is set to $(0,1]$, the model of Eq.(\ref{1}) belongs to
the Ising universality class which has a critical point only at $\lambda
_{c}=1,$ however, when $\gamma =0$, it belongs to the XY universality class
and the critical region is $\lambda _{c}\in (-1,1).$

We assume the central system to be prepared in a superposition state $%
\left\vert \Psi _{S}\right\rangle =\alpha \left\vert 0\right\rangle
+\beta \left\vert 1\right\rangle ,$ thus the initial
system-environment state can be written as $\left\vert \Psi
_{SE}(0)\right\rangle =(\alpha \left\vert 0\right\rangle +\beta
\left\vert 1\right\rangle )\left\vert \Psi _{E}(0)\right\rangle .$
From the evolved reduced density matrix of the system $\rho
_{S}(t)=\mathtt{Tr}_{E}\left\vert \Psi _{SE}(t)\right\rangle
\left\langle \Psi _{SE}(t)\right\vert $, we obtain
\begin{eqnarray}
\rho _{S}(t) &=&\left\vert \alpha \right\vert ^{2}\left\vert
0\right\rangle \left\langle 0\right\vert +\alpha \beta ^{\ast
}R(t)\left\vert 0\right\rangle \left\langle 1\right\vert  \notag \\
&&+\alpha ^{\ast }\beta R^{\ast }(t)\left\vert 1\right\rangle
\left\langle 0\right\vert +\beta ^{2}\left\vert 1\right\rangle
\left\langle 1\right\vert.
\end{eqnarray}%
\label{2}

Clearly, in the basis of the eigenstates $\left\vert 0\right\rangle $ and $%
\left\vert 1\right\rangle $, the diagonal terms in Eq.(2) do not evolve with
time, and only the off-diagonal terms will be modulated by the decoherence
factor $R(t)$, which is the overlap between two states of the environment
obtained by evolving the same initial state $\left\vert \Psi
_{E}(0)\right\rangle $ driven by two different effective Hamiltonians $%
\mathcal{H}_{0}$ and $\mathcal{H}_{1\bigskip }$. As discussed in
\cite{quan}, for the model (\ref{1})\ we have \
$\mathcal{H}_{j\bigskip }=-\sum\nolimits_{l=-M}^{M}\left[
\frac{1+\gamma }{2}\sigma _{l}^{x}\sigma _{l+1}^{x}+\frac{1-\gamma
}{2}\sigma _{l}^{y}\sigma _{l+1}^{y}+(\lambda +j\delta )\sigma
_{l}^{z}\right] $ with $j=0$ or 1. $R(t)$ can be defined as
\begin{equation}
R(t)=\left\langle \Psi _{E}(0)\right\vert e^{it\mathcal{H}_{0}}e^{-it%
\mathcal{H}_{1}}\left\vert \Psi _{E}(0)\right\rangle ,  \label{3}
\end{equation}%
while the LE is determined from $L(t)=\left\vert R(t)\right\vert ^{2},$
which is also called fidelity. If the initial surrounding environment is
prepared in the ground state of $\mathcal{H}_{0}$, i.e., $\left\vert \Phi
_{0}\right\rangle $, Eq.(\ref{3}) will reduce to a simpler form
\begin{equation}
R(t)=\left\langle \Phi _{0}\right\vert e^{-it\mathcal{H}_{1}}\left\vert \Phi
_{0}\right\rangle ,  \label{a}
\end{equation}%
where an irrelevant phase factor is removed.

Next we will deduce the detailed expression of $R(t)$ for the model (\ref{1}%
). In the standard way, the two Hamiltonians $\mathcal{H}_{j}$ can be
diagonalized in terms of a suitable set of fermionic creation and
annihilation operators $\mu _{k}^{(j)}$ as
\begin{equation}
\mathcal{H}_{j}=\sum_{k=1}^{M}\varepsilon _{k}^{(j)}\left[ \mu
_{k}^{(j)\dagger }\mu _{k}^{(j)}-1\right] .  \label{4}
\end{equation}%
When getting the equation above, we have applied to each spin a rotation of $%
\phi $ around the $z$ direction $\mathcal{H}_{j}\left( \phi \right) =g\left(
\phi \right) \mathcal{H}_{j}g^{\dag }\left( \phi \right) $ \ with $g\left(
\phi \right) =\prod\nolimits_{l=-M}^{M}\exp (i\sigma _{l}^{z}\phi /2)$, the
Jordan-Wigner transformation mapping the spins to one-dimensional spinless
fermions with creation and annihilation operators $a_{l}$ and $%
a_{l}^{\dagger }$ via the relation $\ a_{l}=\left( \prod_{i<l}\sigma
_{i}^{z}\right) (\sigma _{l}^{x}+i\sigma _{l}^{y})/2$, and the Fourier
transformation of the fermionic operators described by $c_{k}=\left( 1/\sqrt{%
N}\right) \sum_{l}a_{l}\exp (-i2\pi lk/N)$. The energy spectrum in Eq.(\ref%
{4}) is
\begin{equation}
\varepsilon _{k}^{(j)}=\sqrt{\left[ \cos \left( \frac{2\pi k}{N}\right)
-(\lambda +j\delta )\right] ^{2}+\gamma ^{2}\sin ^{2}\left( \frac{2\pi k}{N}%
\right) },  \label{5}
\end{equation}%
and through a Bogliubov transformation the operators appearing in the
Hamiltonians $\mathcal{H}_{j}$ we have
\begin{equation}
\mu _{k}^{(j)}=c_{k}\cos \left[ \frac{\theta _{k}^{(j)}}{2}\right]
-ic_{-k}^{\dag }e^{2i\phi }\sin \left[ \frac{\theta _{k}^{(j)}}{2}\right] ,
\label{6}
\end{equation}%
where the angles $\theta _{k}^{(j)}$ is the Bogliubov coefficients
satisfying\bigskip\ the equation
\begin{equation}
\cos \left[ \theta _{k}^{(j)}\right] =\frac{\cos \left( \frac{2\pi k}{N}%
\right) -(\lambda +j\delta )}{\varepsilon _{k}^{(j)}}.  \label{7}
\end{equation}%
It is easy to check that the spinless Fermion operators $\mu _{\pm k}^{(j)}$
satisfy
\begin{equation}
\mu _{\pm k}^{(0)}=\mu _{\pm k}^{(1)}\cos \left( \theta _{k}\right) \mp i\mu
_{\mp k}^{(1)\dagger }e^{2i\phi }\sin \left( \theta _{k}\right) ,  \label{8}
\end{equation}%
where $\theta _{k}=$ $\left[ \theta _{k}^{(0)}-\theta _{k}^{(1)}\right] /2.$

According to Eq.(\ref{8}), the ground state of $\mathcal{H}_{0}$ can be
expressed as
\begin{equation}
\left\vert \Phi _{0}\right\rangle _{XY}=\prod\limits_{k=1}^{M}\left[ \cos
(\theta _{k})\left\vert 0\right\rangle _{k}\left\vert 0\right\rangle
_{-k}+ie^{2i\phi }\sin (\theta _{k})\left\vert 1\right\rangle _{k}\left\vert
1\right\rangle _{-k}\right] ,  \label{9}
\end{equation}%
for any opetators $\mu _{\pm k}^{(0)}$ we have $\mu _{\pm k}^{(0)}\left\vert
\Psi _{0}\right\rangle =0$. $\left\vert 0\right\rangle _{k}$ and $\left\vert
1\right\rangle _{k}$ are the vacuum and single excitation of the $k$th mode,
$\mu _{k}^{(1)}$, respectively.

As expected that the ground state of $\mathcal{H}_{0}$ is taken as the
initial surrounding environment state, after some algebraic manipulations
the decoherence factor is obtained
\begin{equation}
R(t)=\prod\limits_{k=1}^{M}R_{k}(t)=\prod\limits_{k=1}^{M}\left[
\sin ^{2}(\theta _{k})+\cos ^{2}(\theta _{k})e^{i2\varepsilon
_{k}^{(1)}t}\right] , \label{10}
\end{equation}
so we can express the LE as
\begin{equation}
L(t)=\left\vert R(t)\right\vert ^{2}=\prod\limits_{k=1}^{M}[1-\sin
^{2}(2\theta_{k}) \sin^{2}(\epsilon_{k}^{(1)}t)].\label{q}
\end{equation}

The term $R_{k}(t)\equiv \sin ^{2}(\theta _{k})+\cos ^{2}(\theta
_{k})e^{i2\varepsilon _{k}^{(1)}t}$ is a decoherence factor for the
$k$-th mode, and its modulus square is always not larger than one.
It is interesting to mention that the Berry phase of the ground
state in the XY model is of sum form for each mode
\cite{carollo,zhu}, while this decoherence factor (\ref{10})\ is of
multiplying form for each mode. Furthermore, Eq.(\ref{10}) is
analogous to the one for non-interacting spin environments\cite{cc}
and Cucchietti generalized Quan's results\cite{dd} .
\begin{figure}[ht]
\begin{center}
\includegraphics[width=0.23\textwidth,height=0.13\textheight]{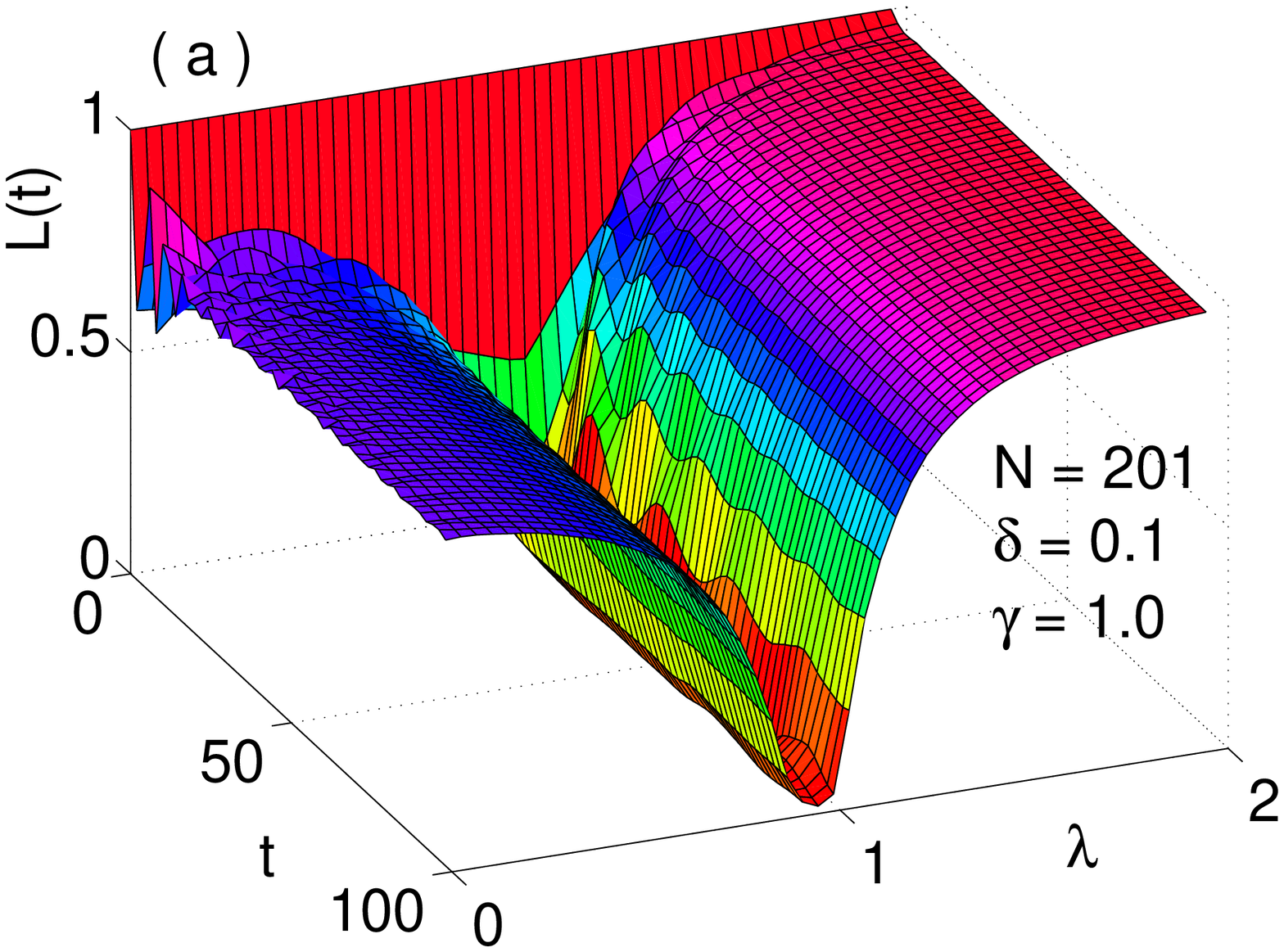} %
\includegraphics[width=0.23\textwidth,height=0.13\textheight]{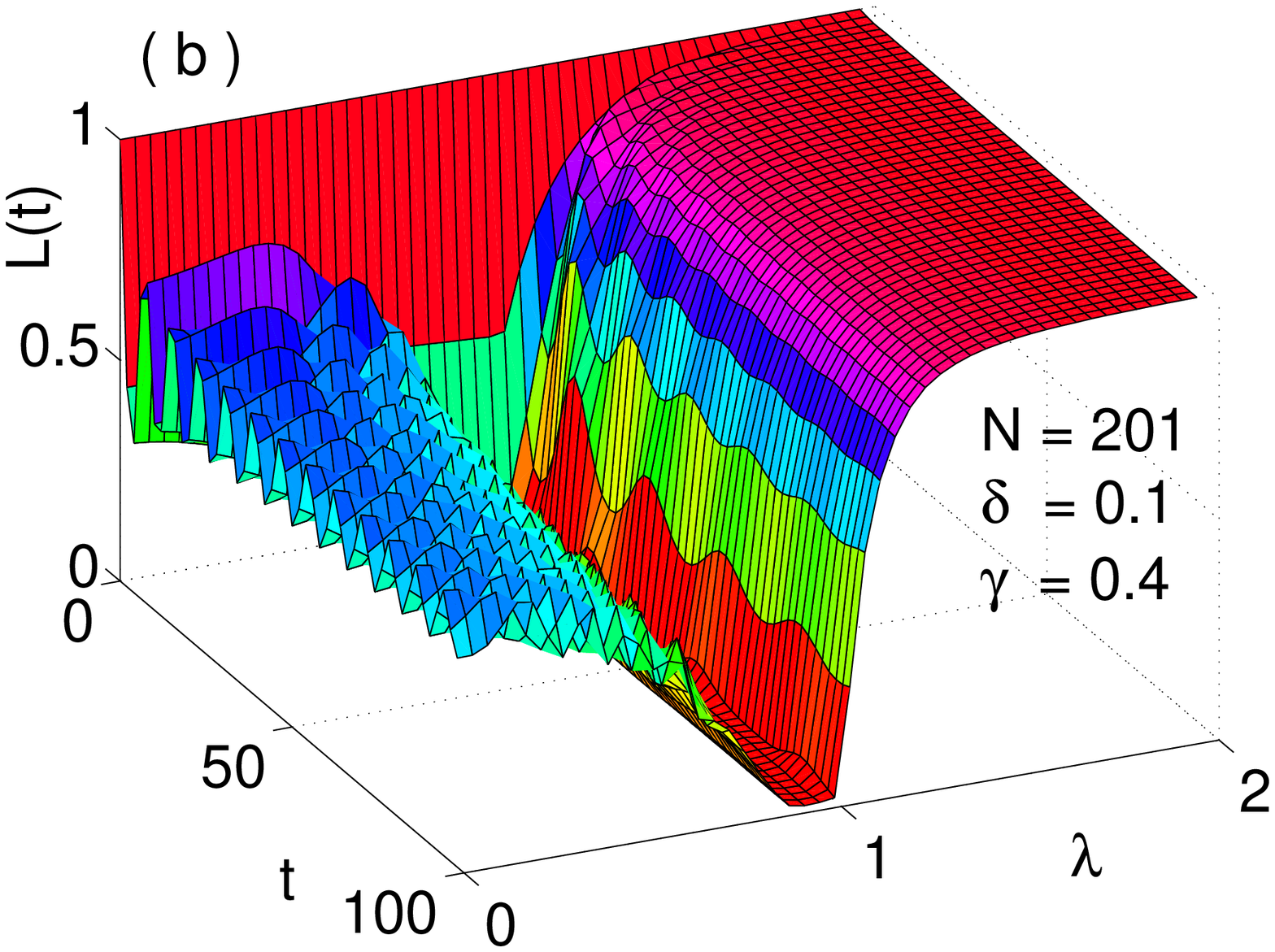} %
\includegraphics[width=0.23\textwidth,height=0.13\textheight]{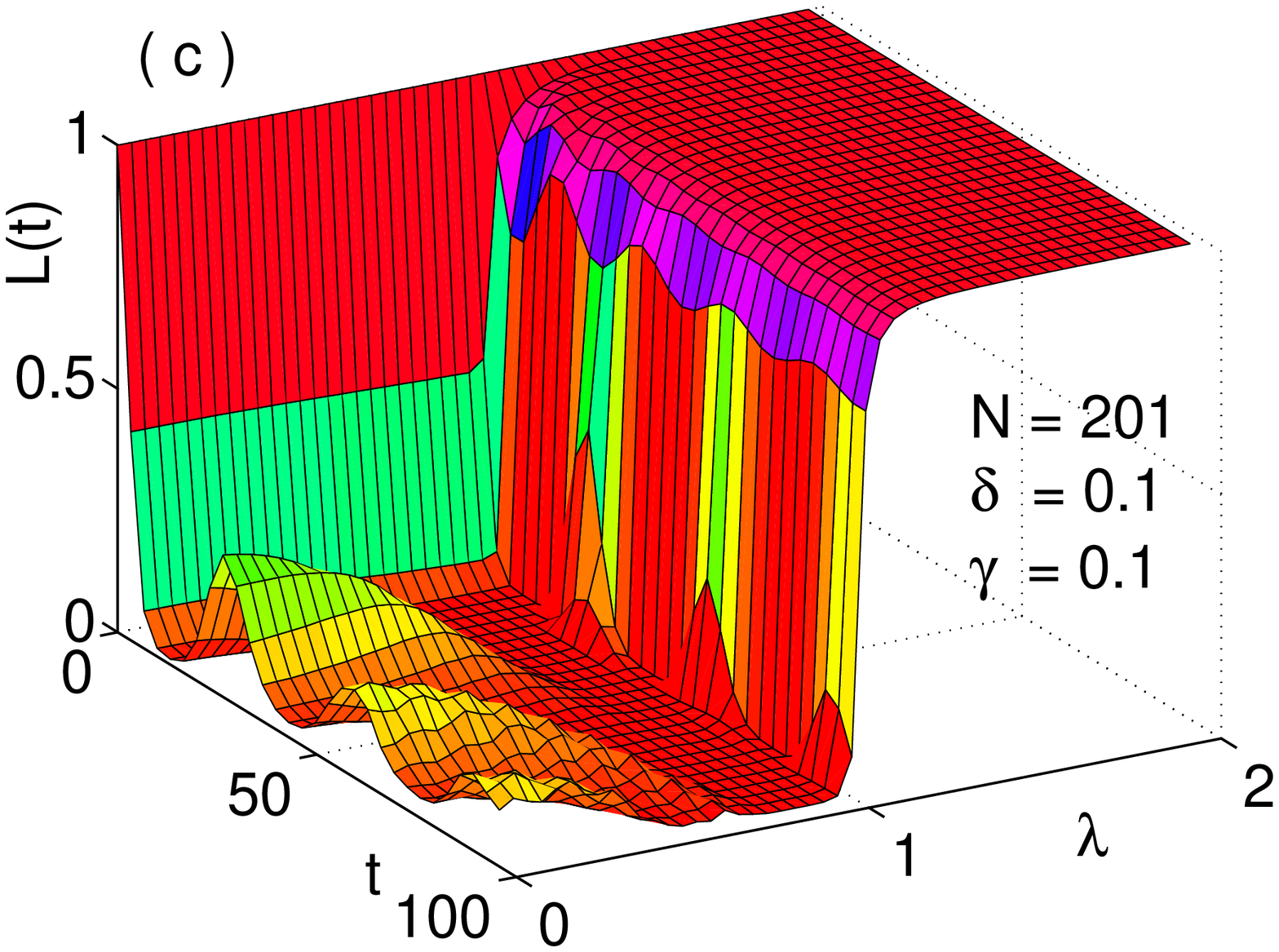} %
\includegraphics[width=0.23\textwidth,height=0.13\textheight]{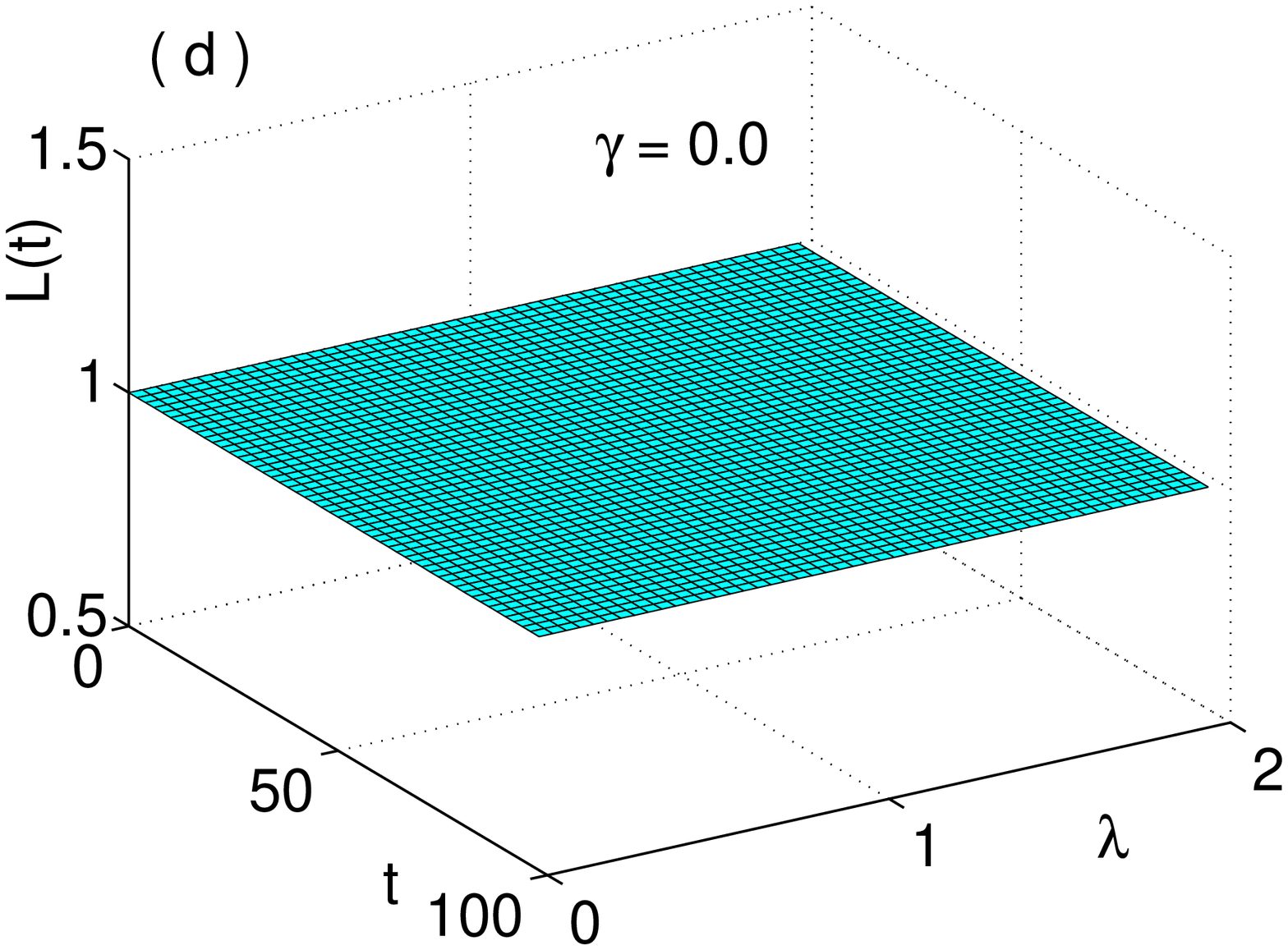}
\end{center}
\caption{ Three dimensional diagram of the $L(t)$ as a function of $t$ and $%
\protect\lambda$ with $N=201$ and $\protect\delta=0.1$. With decreasing $%
\protect\gamma$ from 1.0 to 0.1, (a)-(c) show that the range of $\protect%
\lambda$ where the decay of $L(t)$ is enhanced increases. (d) shows that
when $\protect\gamma=0.0$, the $L(t)$ is unit always, regardless of what $N$%
, $\protect\lambda$ and $\protect\delta$ are.}
\end{figure}

To better understand the LE  (\ref{q}), we plot it as a function of
$t$ and $\lambda$ in
Fig.1, and as a function of only $t$ in Fig.2$.$ For simplicity, we only set $N=201$ and $%
\delta =0.1$. It is demonstrated that the decay of $L(t)$ is enhanced at the
critical point of quantum phase transition $\lambda _{c}=1$ in Fig.1(a),
since the XY model with $\gamma =1$ corresponds to the Ising model. There
exists \ a deep valley in the around the line $\lambda =0.9$, which is the
same results as \cite{quan}. However, for the general XY model where $\gamma
$ is adjusted in $\left( 0,1\right) $, we find that the decay of $L(t)$ is
enhanced in a different degree in the range $\lambda \in (0,1)$. When $%
\gamma =0.4$, the applitudes of $L(t)$ in Fig.1(b) is smaller than
those corresponding to Fig.1(a) and the range of $\lambda $
resulting in $L(t)=0$ increases. When we continue to decrease
$\gamma $ to $0.1$ in Fig.1(c), it is seen that the $L(t)$ nearly
approaches zero in the range $\lambda \in \left( 0,1\right) $, where
the central system transits from a pure state to a mixed state. So
we can conclude that for a smaller $\gamma $, the critical point of
quantum phase transition is the transition point of whether the
decay of the $L(t)$ is enhanced or not.

Comparing with the results in \cite{quan}, for the general XY model,
we also see that $L(t)$ decays and revives as time increases in
Fig.2(e) and (f). This may serve as a witness of QPT.\ At the same
time, if we appropriately adjust the parameters $N$, $\delta $ and
$\lambda $ as shown in Fig.2(e) and (f), it is found that the two
plots of $L(t)$ with the same $\gamma $ have the identical profile,
indicating that the period of the revival of the $L(t)$ is
proportional to the size of the surrounding system in the case of
finite $N$. Fig.2(g) and
(h) reflect that the decreasing $\gamma $ leads to fast decaying of the $%
L(t) $, which complies with the situation described by Fig.1(a)-(c).
In quantum chaos\cite{peres} the sensitivity of perturbations in the
Hamiltonian system can be understood according to the
LE\cite{jalabert}. Here, for some paremeters shown in Fig.2(g) and
(h), the $L(t)$ becomes chaotic, which is due to the competition
between the two phases separated by $\lambda _{c}=1$.

\begin{figure}[tbp]
\includegraphics[width=0.23\textwidth,height=0.13\textheight]{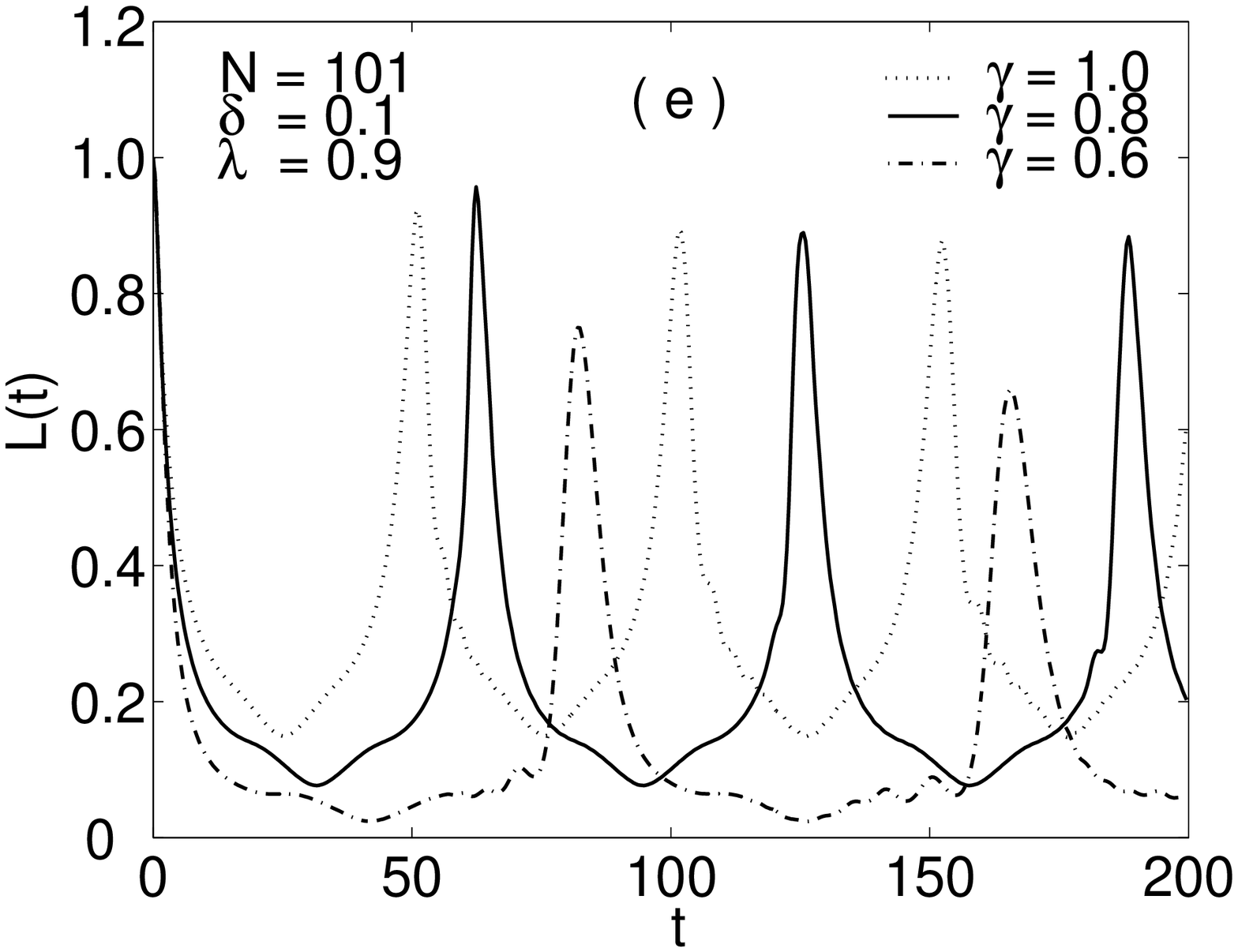} %
\includegraphics[width=0.23\textwidth,height=0.13\textheight]{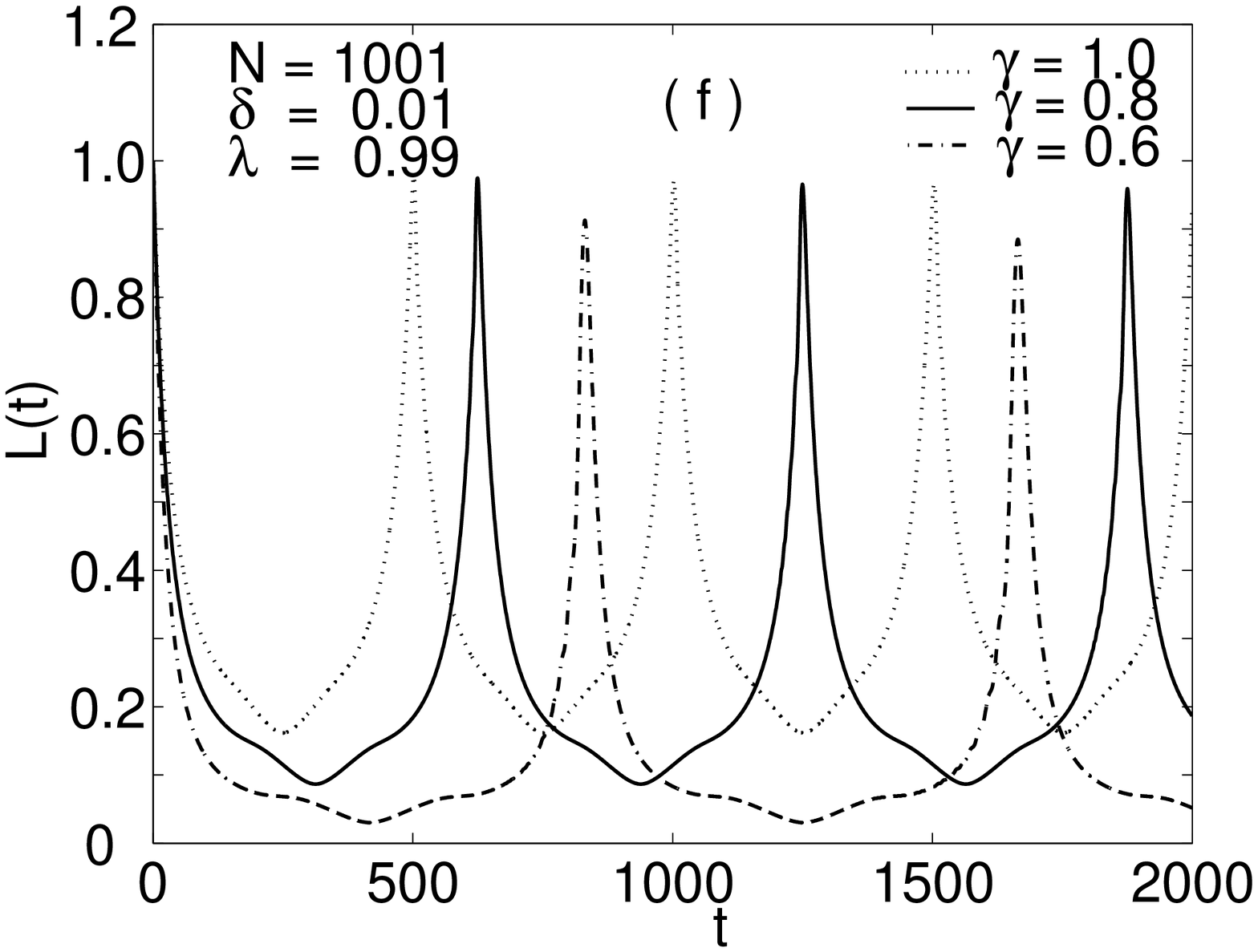} %
\includegraphics[width=0.23\textwidth,height=0.13\textheight]{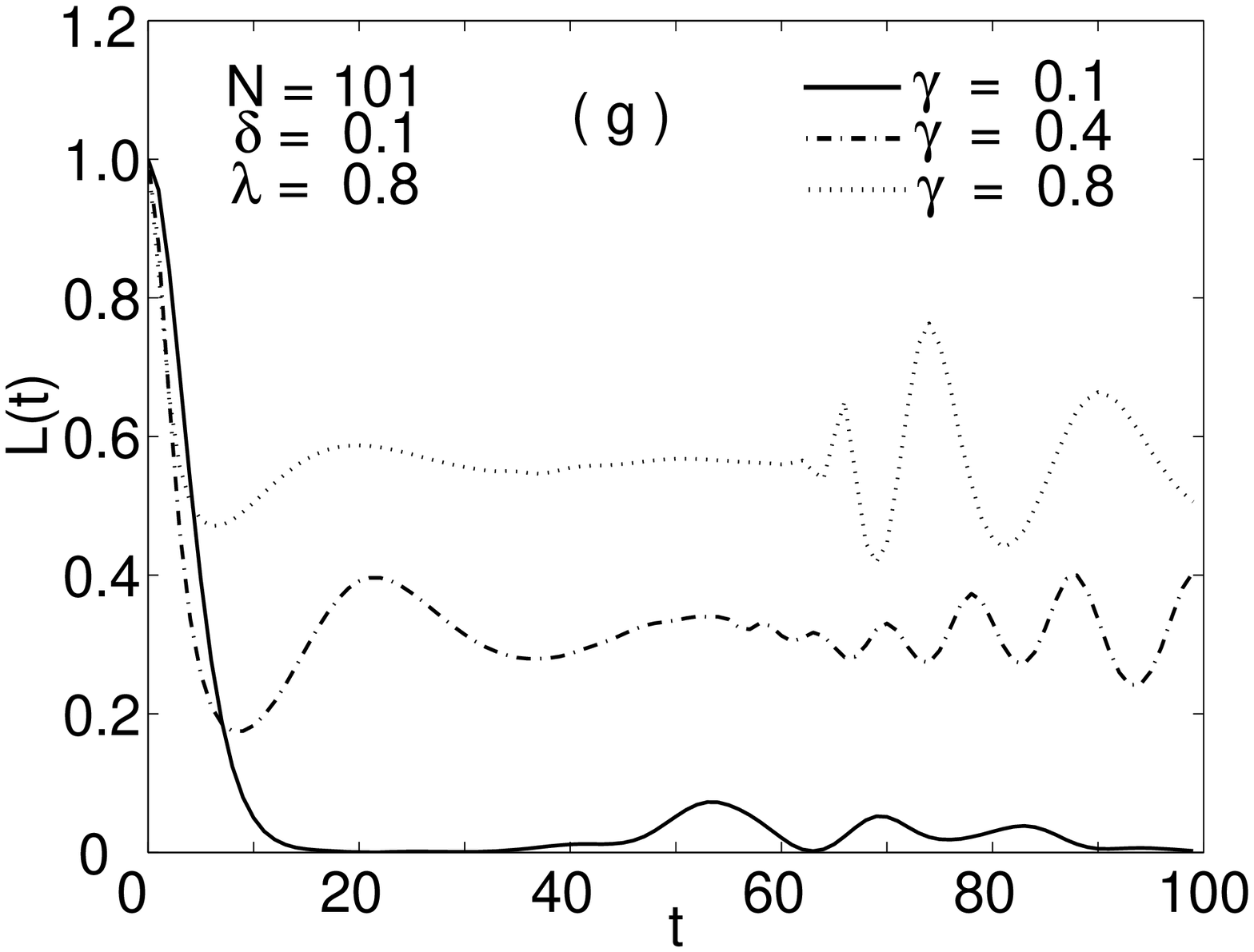} %
\includegraphics[width=0.23\textwidth,height=0.13\textheight]{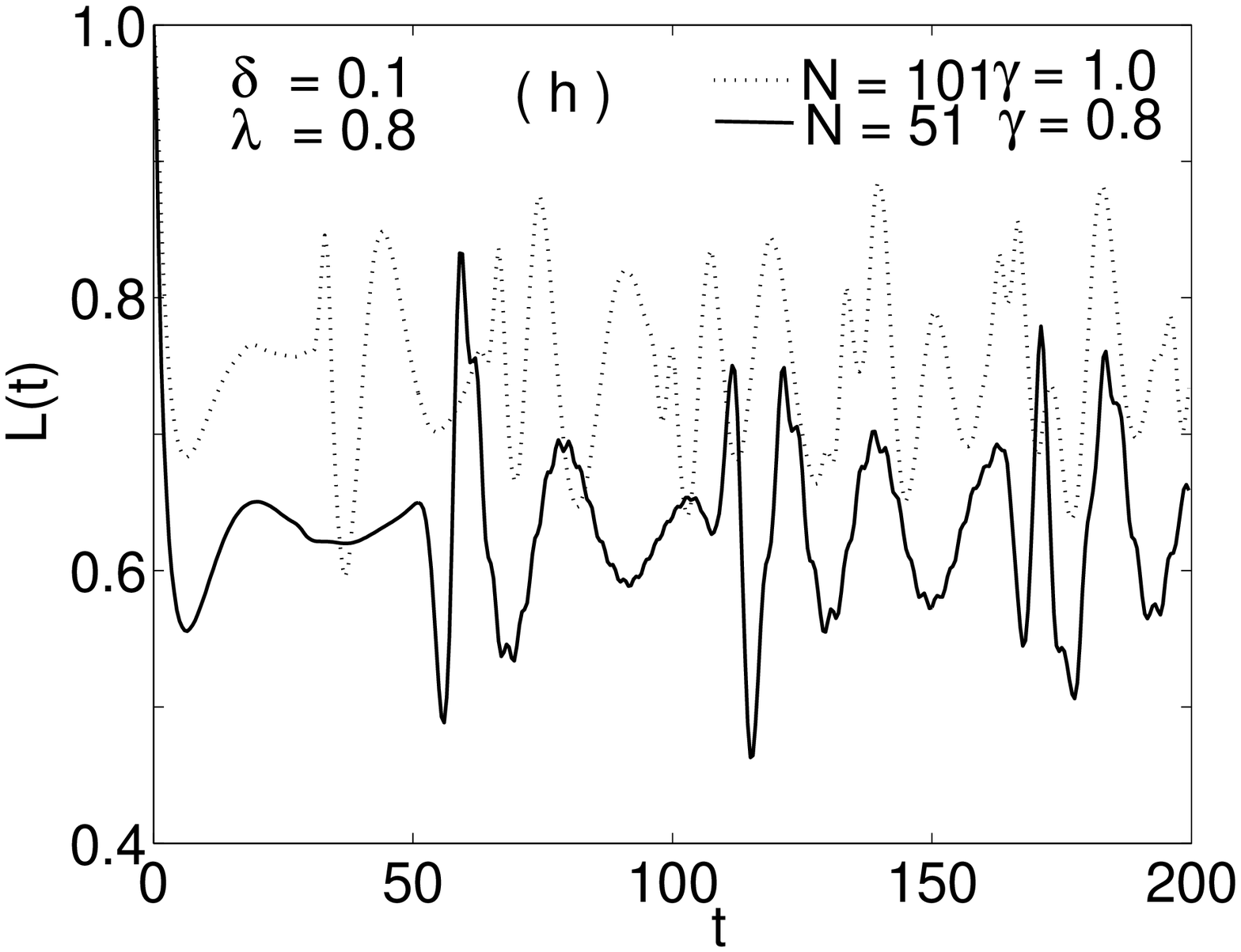}
\caption{The $L(t)$ as a function of $t$. (e) and (f) show that with
decreasing $\protect\gamma$ the quasiperiod of the $L(t)$ increases,
and it is proportional to the size of the surrounding system. The
quasiperiod stems from quantum phase transiton at $\lambda
_{c}=\lambda + \delta$=1. (g) shows that with decreasing
$\protect\gamma$ the decay of the $L(t)$ is
enhanced faster. (h) shows that for some parameters $\protect\lambda$ (away from critical point $\lambda_{c}=1$) the $%
L(t)$ becomes chaotic, which is due to the competition between the
two phases separated by $\lambda _{c}=1$.}
\end{figure}

\section{Analysis of vanishing decoherence}

Interestingly, we find that the $L(t)$ does not vary with time in
the XX model with $\gamma =0$, i.e., the coherence of the central
spin will not be affected by the special environment. The reasons
are in the following. From Eqs.(\ref{5}-\ref{7}), we see that
\begin{equation}
\lim_{\gamma \rightarrow 0}\cos \left[ \theta _{k}^{(j)}\right] =\pm 1,
\label{11}
\end{equation}%
which directly results in that the ground state of $\mathcal{H}_{0}$
no longer lies in the two-dimensional Hilbert space spanned by
$\left\vert 0\right\rangle _{k}\left\vert 0\right\rangle _{-k}$ and
$\left\vert 1\right\rangle _{k}\left\vert 1\right\rangle _{-k}$, but
only one of them like Eq.(\ref{12}). To obtain the explicit form of
the ground state, we let $\cos (2\pi k_{1}/N)=\lambda $ and $\cos
(2\pi k_{0}/N)=\lambda +\delta $ in
Eq.(\ref{7}), and from them know that $k_{0}<k_{1}$. Considering Eqs.(\ref{7}%
-\ref{9}), we can express the ground state
\begin{equation}
\left\vert \Phi _{0}\right\rangle _{XX}=\prod\limits_{k=1}^{k_{0}}\left\vert
0\right\rangle _{k}\left\vert 0\right\rangle
_{-k}\prod\limits_{k=k_{0}+1}^{M}(ie^{2i\phi })\left\vert 1\right\rangle
_{k}\left\vert 1\right\rangle _{-k},  \label{12}
\end{equation}%
since $\cos (\theta _{k})=1$ in Eq.(\ref{9}) when $k<k_{0}$%
, and $\mu _{k}^{(0)}=\mu _{k}^{(1)}$ ($k<k_{0}$) or $-\mu _{k}^{(1)}$ ($%
k>k_{0}$). Therefore the LE becomes
\begin{equation}
L(t)=\lim_{\gamma \rightarrow 0}\left\vert R(t)\right\vert ^{2}=1,
\label{13}
\end{equation}%
which means that the $L(t)$, initial equal to 1, will not decay at
all, and the central system preserves its initial coherence except
for an additional phase factor in one of eigenstate of the central
system. For our model with the central system surrounded by the XX
spin-1/2 bath, notice that the purity \cite{quan,cucchietti} $P=$
Tr$_{S}(\rho _{S}^{2})$ is defined to describe decoherence, it is
$P=1-2\left\vert \alpha \beta \right\vert ^{2}[1-L(t)]=1$ also,
independent of the central system and this special environment.
What's more, the result (\ref{13}) is regardless of the number $N$
of the lattice and what the external magnetic field $\lambda $ is,
it seems to be counterintuitive, but it is indeed the case. We can
see Fig.1(d). At the same time, we emphasize that the non-decay of
the LE in this case is nontrivial because of the difference between
$\mathcal{H}_{0}$ and $\mathcal{H}_{1}$.

The result can be better understood as follows. Evolving from the initial
system-environment state $\left\vert \Psi _{SE}(0)\right\rangle =(\alpha
\left\vert 0\right\rangle +\beta \left\vert 1\right\rangle )\left\vert \Psi
_{E}(0)\right\rangle $, which are not entangled, it becomes $\left\vert \Psi
_{SE}(t)\right\rangle =\alpha \left\vert 0\right\rangle \left\vert \Phi
_{0}(t)\right\rangle +\beta \left\vert 1\right\rangle \left\vert \Phi
_{1}(t)\right\rangle $ at an arbitrary $t$, where $\left\vert \Phi
_{0}(t)\right\rangle $ and $\left\vert \Phi _{1}(t)\right\rangle $ are
driven by the Hamiltonians $\mathcal{H}_{0}$ and $\mathcal{H}_{1}$,
respectively. It is known that $\mathcal{H}_{0}$ and $\mathcal{H}_{1}$ are
different, however, it happens that the anisotropy parameter $\gamma =0$
leads to their same evolution, resulting in $\langle \Phi _{0}(t)\left\vert
\Phi _{1}(t)\right\rangle =\exp (i\varphi )$. The real time-dependent $%
\varphi $ denotes an additional phase factor.

Finally, we assume that the XX model environment to be initially
prepared in an arbitrary excited state with $\gamma =0$. A
$n$-particle state has the form $\mu _{k_{1}}^{(0)\dagger }\mu
_{k_{2}}^{(0)\dagger }\cdot \cdot \cdot \mu _{k_{n}}^{(0)\dagger
}\left\vert \Phi _{0}\right\rangle _{XX} $, with all the $k_{i}$
distinct, i.e., it is
\begin{eqnarray}
\left\vert \Phi _{n}\right\rangle _{XX} &=&\prod\limits_{k^{\prime
}=k_{1}}^{k_{n}}\left\vert 1\right\rangle _{k^{\prime }}\left\vert
0\right\rangle _{-k^{\prime }}\prod\limits_{k=1}^{k_{0}}\left\vert
0\right\rangle _{k}\left\vert 0\right\rangle _{-k}  \notag \\
&&\times \prod\limits_{k=k_{0}+1}^{M}(ie^{2i\phi })\left\vert
1\right\rangle _{k}\left\vert 1\right\rangle _{-k}, \label{ee}
\end{eqnarray}%
where $k\neq k^{\prime }.$ Substituting Eq.(\ref{ee}) into
Eq.(\ref{a}), we find that the LE is $L(t)=1$ also, which implies
that the partial excited states of the environment does not induce
decoherence to the central system. However, if the environment is
initially prepared in a thermal state, the LE is no longer equal to
unit, but will decay with time.

Of particular interest is the case in which the XY model lies
initially in an excited state. The $m$-particle state can be written
as
\begin{eqnarray}
\left\vert \Phi _{m}\right\rangle _{XY} &=&\prod\limits_{k^{\prime
}=k_{1}}^{k_{m}}\left\vert 1\right\rangle _{k^{\prime }}\left\vert
0\right\rangle _{-k^{\prime }}\prod\limits_{k=1,k\neq k^{\prime }}^{M}[\cos
(\theta _{k})\left\vert 0\right\rangle _{k}\left\vert 0\right\rangle _{-k}
\notag \\
&&+ie^{2i\phi }\sin (\theta _{k})\left\vert 1\right\rangle
_{k}\left\vert 1\right\rangle _{-k}]. \label{tt}
\end{eqnarray}%
After calculation by substituting Eq.(\ref{tt}) into Eq.(\ref{a})
the LE\ is derived as $L(t)=\ \prod\nolimits_{k=1}^{M}\left[ 1-\sin
^{2}(2\theta _{k})\sin ^{2}(\varepsilon _{k}^{(1)}t)\right] $ with
$k\neq k^{\prime }$. It can be seen that: (i) the excited states
$\prod\nolimits_{k^{\prime }=k_{1}}^{k_{m}}\left\vert 1\right\rangle
_{k^{\prime }}\left\vert 0\right\rangle _{-k^{\prime }}$ have no any
contributions to modulating the LE;\ \ (ii) if all the particles are
excited, i.e., $\left\vert \Phi _{m}\right\rangle
_{XY}=\prod\nolimits_{k=1}^{M}\left\vert 1\right\rangle
_{k}\left\vert 0\right\rangle _{-k}$, which is assumed to be the
initial state of the bath, the LE of the central system is unit
also. Notice that if the initial state of the XY model bath is
thermal, the LE of the central system will decay with time. The
vanishing decoherence may arise in view of the non-interacting
environments in \cite{cc}: if we let all spins lie initially in
either up or down, the decoherence factor will be unit as well. It
is worthwhile for us to find out its physical nature.

\section{Conclusions}

In summary, for a central system uniformly coupled to a XY spin-1/2
bath in a transverse field, we explicitly calculate the Loschmidt
echo(LE) to characterize decoherence quantitatively. We find that
the anisotropy parameter $\gamma$
affects decoherence of the central system sensitively when $\gamma \in [0,1]$%
, in particular, the LE becomes unit without varying with time when
$\gamma =0 $ , implying that environment-induced decoherence
vanishes. At the same time, we show that decoherence can vanish
provided that the initial state of the XX spin-1/2 bath lies in
either the ground state or the partial particles are excited, or it
lies in the state that all particles are excited. Although it is
difficult to make the initial state of the spin bath pure at zero
temperature and then difficult to fulfil the vanishing decoherence
in real experiments, in a theoretical sense, our findings may shed
light on understanding of decoherence.

\bigskip

\section{Acknowledgement}

We greatly acknowledge the helpful discussions with H.T.Quan and X.Z.Yuan.


\begin{references}
\bibitem{AGG} W. G .Unruh, {Phys. Rev. A } {\bf 51}, 992 (1995).
\bibitem{A1GG} D. P. DiVincenzo, Science {\bf 270}, 225 (1995).
\bibitem{CS} A. R. Calderbank and P. W. Shor {\bf 54}, 1098 (1996).
\bibitem{S} A. M. Steane, Phys. Rev. Lett.  {\bf 77}, 793 (1996).
\bibitem{Gottesman} D. Gottesman, \emph{Stabilizer Codes and Quantum Error Correction}. Ph.D. thesis, California Institute of
Technology, Pasadema, CA, 1997.
\bibitem{Duan} D. Bacon, J. Kempe, D. A. Lidar, and K. B. Whaley, Phys. Rev. Lett.  {\bf 85}, 1758 (2000).
\bibitem{Lidar} L. M. Duan and G. C. Guo, Phys. Rev. Lett.  {\bf 79}, 1953 (1997).
\bibitem{zurek} W. H. Zurek, Rev. Mod.Phys. {\bm 75}, 715 (2003); Phys. Today {\bf 44}, No. 10, 36(1991).
\bibitem{peres} A. Peres, \emph{Quantum Theory: Concepts and Methods},
(Kluwer Academic Publishers, Dordrecht, 1995).
\bibitem{weiss} U. Weiss, \textit {Quantum Dissipative Systems}, 2nd
ed. (World Scientific, Singapore, 1999).
\bibitem{osborne} T. J. Osborne and M. A. Nielsen, Phys. Rev. A  {\bf 66}, 032110 (2002).
\bibitem{vidal} L. A. Wu, M. S. Sarandy, and D. A. Lidar, Phys. Rev. Lett.  {\bf 93}, 250404 (2004).
\bibitem{carollo} C. M. Carollo and J. K. Pachos, Phys. Rev. Lett.  {\bf 95}, 157203 (2005).
\bibitem{zhu} S. L. Zhu, Phys. Rev. Lett.  {\bf 96}, 077206 (2006).
\bibitem{hamma} A. Hamma, quant-ph/0602091.
\bibitem{quan} H. T. Quan, Z. Song, X. F. Liu, P. Zanardi, and C. P. Sun, Phys. Rev. Lett.  {\bf 96}, 140604 (2006).
\bibitem{rossini} D. Rossini, T. Calarco, V. Giovannetti, S.Montangero, R. Fazio, quant-ph/0605051.
\bibitem{book} S. Sachdev, \textit{Quantum Phase Transition} (Cambridge University Press, Cambridge, 2000).
\bibitem{cucchietti} F. M. Cucchitti, D. A. R. Dalvit, J. P. Paz, and W. H. Zurek, Phys. Rev. Lett. {\bf 91}, 210403(2003).
\bibitem{cc}  F. M. Cucchitti, J. P. Paz, and W. H. Zurek, Phys. Rev. A {\bf 72}, 052113(2005).
\bibitem{dd}F. M. Cucchitti, S. F. Vidal, and J. P. Paz, quant-ph/0604136.
\bibitem{jalabert} R. A. Jalabert and H.M. Pastawski, Phys. Rev. Lett.
{\bf 86}, 2490 (2001).
\end{references}
\end{document}